\documentclass[aps,prb,showpacs,twocolumn,superscriptaddress,nobibnotes]{revtex4-1}

\RequirePackage{subfigure}
\usepackage{lipsum}
\usepackage{graphicx}
\usepackage{color}
\usepackage{enumerate}
\usepackage{amsmath}
\usepackage{amssymb}
\usepackage{stmaryrd}
\usepackage{multirow}
\usepackage[normalem]{ulem}
\usepackage{float}
\usepackage{longtable,booktabs}
\usepackage[dvipsnames,table,xcdraw]{xcolor}
\usepackage[normalem]{ulem}
\usepackage{physics}
\usepackage{url}
\usepackage{soul} 
\usepackage[colorinlistoftodos]{todonotes}
\definecolor{darkblue}{rgb}{0.1,0.2,0.6} \definecolor{darkred}{rgb}{0.8,0.1,0.2}
\usepackage[colorlinks,citecolor=darkblue,linkcolor=darkred,urlcolor=darkblue]{hyperref}
\usepackage[all]{hypcap} 

\usepackage[draft]{pdfcomment}

\def\beq{\begin{equation}}
\def\eeq{\end{equation}}

\begin{document}
\long\def\/*#1*/{}
\title{Beyond many-body localized states in a spin-disordered Hubbard model with pseudo-spin symmetry}

\author{Xiongjie Yu}
\thanks{These authors contributed equally to this work.}
\affiliation{Institute for Condensed Matter Theory and Department of Physics, University of Illinois at Urbana-Champaign, IL 61801, USA} 

\author{Di Luo}
\thanks{These authors contributed equally to this work.}
\affiliation{Institute for Condensed Matter Theory and Department of Physics, University of Illinois at Urbana-Champaign, IL 61801, USA} 

\author{Bryan K.  Clark}
\affiliation{Institute for Condensed Matter Theory and Department of Physics, University of Illinois at Urbana-Champaign, IL 61801, USA} 

\begin{abstract}
A prime characterization of many-body localized (MBL) systems is the entanglement of their eigenstates; in contrast to the typical ergodic phase whose eigenstates are volume law, MBL eigenstates obey an area law.  In this work, we show that a spin-disordered Hubbard model has both a large number of area-law eigenstates as well as a large number of eigenstates whose entanglement scales logarithmically with system size (log-law). This model, then, is a microscopic Hamiltonian which is neither ergodic nor many-body localized. We establish these results through a combination of analytic arguments based on the eta-pairing operators \cite{Vafek_2016arXiv_eta} combined with a numerical analysis of eigenstates. In addition, we describe and simulate a dynamic time evolution approach starting from product states through which one can separately probe the area law and log-law eigenstates in this system. 
\end{abstract}
 
\maketitle

\section{Introduction}
Pioneered by the seminal works of Basko \cite{basko_metalinsulator_2006} and Gornyi \cite{gornyi_interacting_2005},
the many-body localization (MBL) transition is defined as a dynamical phase transition which happens at finite energy density for a disordered and isolated many-body interacting system. Conceptually, MBL is when Anderson localization \cite{anderson_absence_1958,Fleishman_Interactions_1980} survives inter-particle interactions. In MBL systems, under unitary time evolution, local observables fail to thermalize to their ergodic values. 

Typical MBL models are disordered spin chains with short-ranged interactions in one dimension \cite{oganesyan_localization_2007,znidaric_many-body_2008,Berkelbach_Conductivity_2010,pal_many-body_2010,kjall_many-body_2014,luca_ergodicity_2013,pekker_encoding_2014,Bar_Lev_Dynamics_2014,Bar_Lev_Absence_2015,Agarwal_Anomalous_2015,luitz_many-body_2015,yu_finding_2015,Yu2016_bimodal,serbyn_criterion_2015,bera_many-body_2015,luitz_extended_2016,singh_signatures_2016,pollmann_efficient_2015,khemani_obtaining_2015,lim_nature_2015,luitz_long_2016,luitz_anomalous_2016,bera_local_2016,serbyn_universal_2016,Khemani_Critical_2017,Luitz_Information_2017,Imbrie_Diagonalization_2016}.
More recently, systems with itinerant degrees of freedom have been explored including disordered Hubbard or t-J models \cite{Prelovsek_2016PRB_absence,Jakub_2018arXiv,Bar_Lev_delocalized,Lemut_Complete_2017}. This focus has been partially motivated by cold-atom experiments \cite{Schreiber2015, Kondov2015, Bordia_2016PRL_Coupling,Luschen_Evidence_2016}. 

Phenomenologically, the full MBL (FMBL) phase is characterized by a complete set of local integrals of motion (LIOM)  \cite{Joel_unbound_2012,serbyn_local_2013,imbrie_many-body_2014,huse_phenomenology_2014,chandran_constructing_2015,Chandran2014,ros_integrals_2015,
Pekker_Fixed_2017,inglis_accessing_2016,pekker_encoding_2014,Wahl2016,Imbrie_Review_2017,
monthus_many-body_2016}, or the existence of a small bond-dimension unitary tensor network (UTN) which diagonalizes the MBL Hamiltonian \cite{pekker_encoding_2014,Vidal_spectral_2015}. A key application of the LIOMs or UTN is to explain the entanglement behavior of the MBL system. They imply that the entanglement of eigenstates are area law and that entanglement grows logarithmically under time evolution after a quench \cite{Issac_lightcone}.

In this work, we report on a microscopic Hamiltonian which goes beyond the FMBL or ergodic phases.  We show that this microscopic Hamiltonian has both constant (area law) as well as logarithmically entangled (log law) eigenstates. These eigenstates are interspersed throughout the spectrum (i.e. they don't make up a mobility edge).  We then show how to probe separately the area-law and log-law eigenstates through time-evolution from simple product states giving potential access to these different types of states through cold-atom experiments.

\begin{figure}[H]
  \centering
  \includegraphics[scale=0.6]{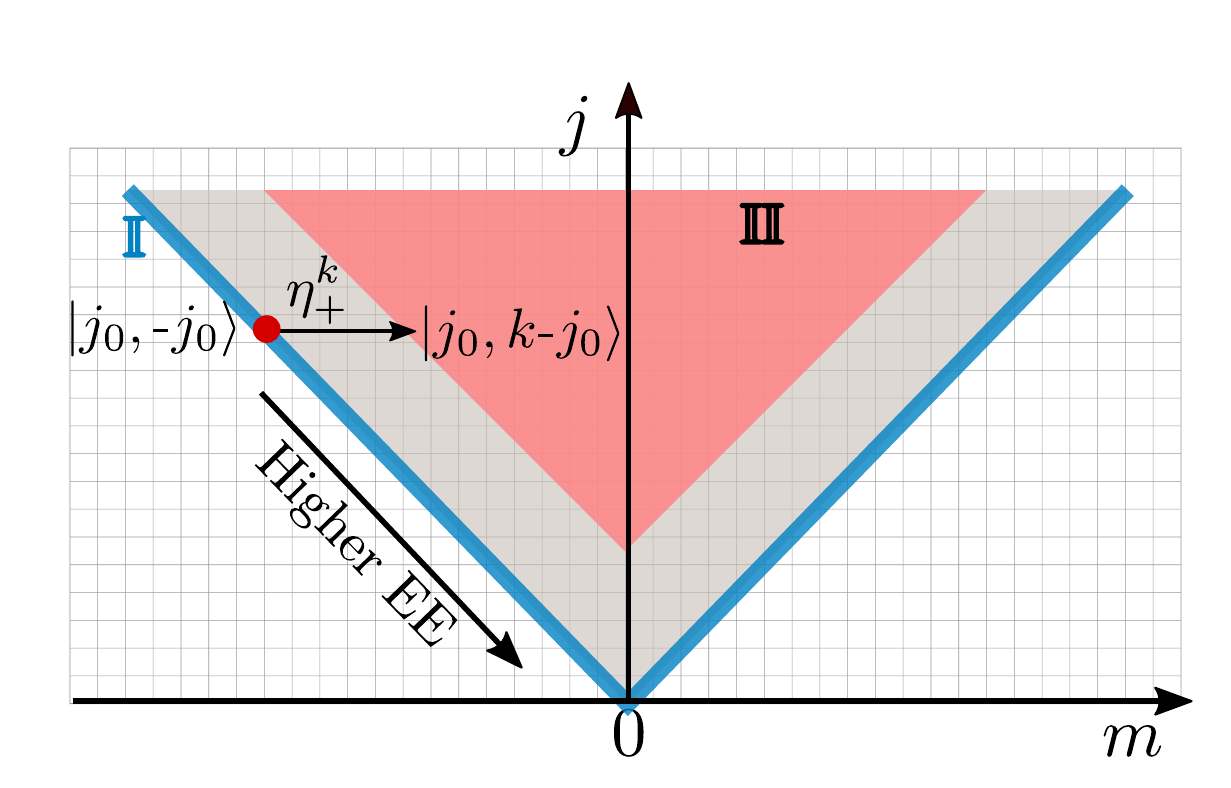}
	\caption{Schematic phase diagram of the spin-disordered Hubbard chain at large disorder in the the 2D plane labeled by quantum numbers $j$ and $m$. 
The eta-pairing raising operator $\eta_+$  acts as $\eta_+ |E, S^z_\textrm{total},j,m\rangle = |E+U, S^z_\textrm{total},j,m+1\rangle$ moving eigenstates horizontally in this figure. The particle-hole transformation maps $|E, S^z_\textrm{total}, j, m \rangle$ to $|E-2mU, S^z_\textrm{total}, j, -m \rangle$, equivalent to a mirror symmetry about $m=0$. In the $S^z_\textrm{total} = 0$ sector, the top left corner of the triangle is the vacuum state. \textbf{Region I} 
(\textcolor{blue}{blue}) are reference states which are destroyed by $\eta_-$ (left edge) or $\eta_+$ (right edge) and contain a mixture of area-law and log-law states.  
\textbf{Region II} (\textcolor{pink}{pink}) is the region where all eigenstates' entanglement entropies have logarithmic correction due to repeated application of $\eta_+$ (left edge) or $\eta_-$ (right edge). 
}
  \label{fig:EE_Hmap}
\end{figure}

\subsection{Overview of Results}
We consider a one-dimensional Hubbard model with spin disorder
\begin{equation}
H = -t \sum_{i\sigma} (c^{\dagger}_{i\sigma} c_{i+1\sigma} + h.c.) +  \sum_i U n_{i\uparrow} n_{i\downarrow} + \sum_i h_i S^{z}_{i},\label{eq:HB_spin}
\end{equation}
where $S^{z}_{i}=(n_{i\uparrow} - n_{i\downarrow})/2$, and the disordered magnetic field $h_i \in [-W,W]$ is sampled uniformly. We focus on the case of $U=1, t=1$.  

To scaffold this discussion, we first note that Eq.~\eqref{eq:HB_spin} has a pseudo-spin SU(2) symmetry \cite{Yang_1989PRL_eta,Zhang_1990PRL_eta} (see Sec.~\ref{sec:Model}), which allows us to label our eigenstates by four quantum numbers $|E,S^z_\textrm{total},j,m\rangle$ including the energy $E$, total $S_z$ and two quantum numbers $j$ and $m$ associated with the pseudo-spin symmetry.   Because this symmetry is a continuous non-abelian symmetry, we don't expect to have a fully-MBL phase \cite{Potter_Symmetry_2016,Protopopov_Effect_2017}.  Unless otherwise noted, we work with $S_\textrm{total}^z=0$ (although our results generically apply to all $S_\textrm{total}^z$) and separately consider the entanglement of eigenstates in different quantum number sectors.  Fig.~\ref{fig:EE_Hmap} is a diagram of available quantum numbers and this figure will set the framework in which we discuss our results. 

From the pseudo-spin algebra, one can analytically build towers of excited states of increasing $m$ using the eta-pairing raising and lower operators $\eta_+$ and $\eta_-$  \cite{Yang_1989PRL_eta,Zhang_1990PRL_eta};  every application of $\eta_+$ moves states horizontally in Fig.~\ref{fig:EE_Hmap}.  The blue line (region I) are the eigenstates at the bottom of the towers which we call reference states.  We  then consider eigenstates in region II which are generated from a reference state by the application of $\eta_+^N$ for a given constant $N$.  A common feature of this group of excited states is the large number of double occupancies.  We will show that all such eigenstates have, at least, an additive logarithmic correction to their von Neumann entanglement entropy with respect to the reference state; this violates the area-law entanglement for a typical MBL phase.    We accomplish this by identifying a particular sector of the reduced density matrix for these states that leads to the logarithmic correction  (see Sec.~\ref{sec:EE_corrections}).  This extends results of Ref.~\onlinecite{Vafek_2016arXiv_eta} which recently showed such corrections in the case of the vacuum state (top left of Fig.~\ref{fig:EE_Hmap}) with applications to a non-disordered Hubbard model.  Additionally, we show that any eigenstate which is made of only singlons has an exact logarithmic correction. 

We then numerically consider a number of disorder realizations of the Hamiltonian in Eq.~\eqref{eq:HB_spin} for $L=8$ at large $W$ (see Sec.~\ref{sec:Entanglement_of_reference_states}) using the slope of the cut average entanglement (SCAEE), introduced in Ref.~\onlinecite{Yu2016_bimodal}.  We find that the reference eigenstates contain a mixture of area-law and log-law states and that the full spectrum of eigenstates in region II do indeed exhibit a logarithmic increase in entanglement  (see Fig.~\ref{fig:dEE_lnv_1-v}).

Having characterized the eigenstates, we then discuss how to separately probe the localization physics both from area-law states as well as from log-law states dynamically using time evolution (see Sec. \ref{sec:Time_Evolve}). This would allow cold-atom experiments to directly probe this physics.  We identify two extreme cases of product states --
all single occupancies at quarter filling which occupy primarily area law states, and all double occupancies at half filling which occupy  log-law states. We find that in former case the entanglement entropy grows logarithmically and the charge imbalance does not relax as is typical in a
MBL system, while in the latter the entanglement entropy grows as a power law (but not linear) fashion and the charge imbalance tends to fully relax, which is  delocalized but not ergodic.

\section{Introduction to Pseudo-spin algebra} \label{sec:Model}
\subsection{Pseudo-spin $SU(2)$ symmetry}
It is easy to see that the spin disorder breaks the spin rotation symmetry of $H$.
To prove that the pseudo-spin $SU(2)$ symmetry is intact,
one can introduce the eta-pairing operators \cite{Yang_1989PRL_eta,Zhang_1990PRL_eta},
with notations for 1D specifically. 
\begin{equation}
\eta_- = \sum_{i} (-1)^i c_{i\uparrow} c_{i\downarrow}, \ \eta_+ = \eta^{\dagger}_-, \ \eta_0 = \frac{1}{2}(\hat{N}-L),
\end{equation}
where $L$ is the number of sites and has to be even,
and $\hat{N}$ is the operator for total number of electrons in the system.
The eta-pairing operators generate a $SU(2)$ algebra because
\begin{equation}\label{eq:eta_algebra}
[\eta_0, \eta_{\pm}] = \pm \eta_{\pm}, \quad [\eta_+, \eta_-] = 2\eta_0.
\end{equation}

To prove that pseudo-spin symmetry is preserved, one can straightforwardly check that
\begin{equation}\label{eq:H_eta_commutation}
[H, \eta_{\pm}] = \pm U \eta_{\pm}, \ [H, \eta_0] = 0, \ [H, \vec{\eta}^2]=0,
\end{equation}
where the total pseudo-spin operator $\vec{\eta}^2$ is
\begin{equation}
  \vec{\eta}^2 = \frac{1}{2}(\eta_+ \eta_- + \eta_- \eta_+) + \eta^2_0.
\end{equation}
Therefore, $\{H, S^z_\textrm{total}, \vec{\eta}^2, \eta_0 \}$ is a complete set of commuting observables.
For simplicity, we will denote the eigenstate of $\{ \vec{\eta}^2, \eta_0 \}$ as $|j, m\rangle$, with
\begin{equation}
\vec{\eta}^2 |j, m\rangle = j(j+1) |j, m\rangle, \quad \eta_0 |j, m\rangle = m |j, m\rangle,
\end{equation}
where $|m| \le j$.

Because of Eq.~\eqref{eq:eta_algebra} and \eqref{eq:H_eta_commutation}, 
$\eta_{\pm}$ are a pair of ladder operators for $\eta_0$ and the Hamiltonian $H$.
Consider a simultaneous eigenstate $|E, S^z_\textrm{total}, j, m \rangle$. Applying  $\eta_+$ to this eigenstate will lead to 
$|E+U, S^z_\textrm{total}, j, m+1 \rangle $, and vice versa for $\eta_-$.  

\subsection{Particle-hole transformation}
The particle-hole (PH) transformation
\begin{equation}\label{eq:PHT}
c_{i\uparrow} \to (-1)^i c^{\dagger}_{i\downarrow}, \quad c_{i\downarrow} \to (-1)^i c^{\dagger}_{i\uparrow}
\end{equation}
has some consequences for the eigenstates. 

Firstly, $S^z_i$ is invariant under the PH transformation. Secondly, we have
\begin{equation}
\eta_\pm \to \eta_{\mp}, \ \eta_0 \to -\eta_0, \ \vec{\eta}^2 \to \vec{\eta}^2.
\end{equation}
Thirdly, the Hamiltonian $H$ transforms as
\begin{equation}
H \to H  - 2 U\eta_0.
\end{equation}
Therefore, from the wave function's perspective, an eigenstate $|E, S^z_\textrm{total}, j, m \rangle$ under the PH transformation becomes $|E-2mU, S^z_\textrm{total}, j, -m \rangle$.
Both these eigenstates will have the same entanglement; therefore, both the right and left edge of Fig.~\ref{fig:EE_Hmap} can be considered reference states.

\section{Entanglement entropy of eta-pairing states} \label{sec:EE_corrections}
The pseudo-spin $SU(2)$ algebra has direct influence on the eigenstates' energy and entanglement entropy. 
Given a reference eigenstate $|\psi_{\text{ref}}\rangle$ of $H$, 
one can build a tower of highly excited states with $\eta_+$
\begin{equation}
  |\psi^N\rangle = \mathcal{A}_N \eta_+^N |\psi_{\text{ref}}\rangle, \quad N \in \mathbb{N}^+,
\end{equation}
where $\mathcal{A}_N$ is the normalization factor. 
With increasing $N$, $|\psi^N\rangle$ has increasing energy and number of doublons, until annihilated.
We will call these excited states the eta-pairing states, and only consider the reference state $|\psi_\text{ref}\rangle$ which can be annihilated by $\eta_-$ and has relatively small number of electrons (see Appendix \ref{app:reference_state} for details).

In this section, we prove two things:  (A) eta-pairing states (with large enough $N$) have, at least, a logarithmically increasing entanglement with respect to its reference state and (B) eta-pairing states (with large enough $N$) whose reference state consist of only singlons have exactly a logarithmically increasing entanglement. 

To accomplish this, we decompose $|\psi_{ref}\rangle = \sum_t |\psi_t\rangle $ into a linear superposition of terms labeled by property $t$ which is preserved under the application of $\eta_+^N$.  Additionally, the reduced density matrix is block diagonal in blocks labeled by $t$, i.e. $\rho_t = tr_B |\psi_t\rangle\langle \psi_t|$.  It then follows that the entanglement entropy is a sum of these individual blocks. To determine the change of entanglement, we need consider only how the entanglement of each term $|\psi_t\rangle$ changes. 

To prove (A), the property $t$ is the spin polarization in the subsystem $A$.  We consider only the term where $S_{A,z}$ is maximally polarized (i.e. $S_{A,z}=K/2$ for a system of $K$ electrons) and show that this term has a logarithmically increasing entanglement.  To prove (B), the property $t$ is the singlon number in subsystem $A$ and we can show that every term has a logarithmically increasing entanglement.   Proving the logarithmic increase in entanglement uses a similar approach to Ref. \onlinecite{Vafek_2016arXiv_eta}.   In the subsections below we detail these claims.

\subsection{Maximally Polarized Sector}\label{subsec:max_polarization}
In this subsection, we show that, for any eta-pairing eigenstate in region II of Fig. \ref{fig:EE_Hmap}, the entanglement entropy grows at least logarithmically, whose contribution comes from the maximally polarized sector in the reduced density matrix.

Consider an eta-pairing state built on a many-body reference state with $K$ electrons. Without loss of generality, let $S_z=0$. Decompose $|\psi_{ref}\rangle$ into terms of fixed $S_{A,z}$.  Notice that the operation of $\eta_+$ only adds doublons to a basis vector and therefore, can't change the value of $S_{A,z}$ except by destroying the state. Since $S_z$ is fixed, $S_{B,z}$ will not change either when $\eta_+$ is applied. When we trace out $B$ to calculate the reduced density matrix of $A$, the terms $|\psi_{t'}\rangle\langle \psi_t|$ where $t\neq t'$ will vanish because $|\psi_{t'}\rangle$ and $| \psi_t\rangle $ have different values $S_{B,z}$. As a result, the reduced density matrix will be block diagonal according to $S_{A,z}$.

Take a reference state with $K$ electrons for the disordered Hubbard model.
\begin{equation}\label{many_wf}
  |\psi_{\text{ref}}\rangle = \sum_{(i_1,\sigma_1), \cdots, (i_K,\sigma_K)} \alpha_{(i_1,\sigma_1), \cdots, (i_K,\sigma_K)} c^{\dagger}_{i_1,\sigma_1} \cdots c^{\dagger}_{i_K,\sigma_K} |0\rangle,
\end{equation}
which satisfies $K \ll L/2$ and $\eta_- |\psi_{\text{ref}}\rangle = 0$.
We consider the block in the reduced density matrix with maximum $S_{A,z}$ for which there is $K/2$ spin-up electrons in $A$ and $K/2$ spin-down electrons in $B$.

\begin{align}
	|\psi_{K/2} \rangle &= \sum_{(i_1,\uparrow), \cdots, (i_K,\downarrow)} \alpha_{(i_1,\uparrow), \cdots, (i_K,\downarrow)} c^{\dagger}_{i_1,\uparrow},\cdots, c^{\dagger}_{i_K,\downarrow} |0\rangle \\
 & = \sum_{i \in I,j \in J} \alpha_{i,j} \{c^{\dagger}_{i \uparrow}\} \{c^{\dagger}_{j \downarrow}\} |0\rangle.
\end{align}
$I$ is the set of site sequences with only spin-up electrons. $\{c^{\dagger}_{i \uparrow}\}$ is the product of $c^{\dagger}_{\uparrow}$ from a particular site sequence $i$. Similar notation is used for $J$ and $\{c^{\dagger}_{j \downarrow}\}$ for the case of spin-down electrons. We then perform a Schmidt decomposition on $|\psi_{K/2} \rangle$.

\begin{align}
|\psi_{K/2} \rangle &= \sum_{i \in I,j \in J} \alpha_{i,j} \{c^{\dagger}_{i \uparrow}\} \{c^{\dagger}_{j \downarrow}\} |0\rangle\\
&= \sum_{i \in I,j \in J} \sum_{k} u_{i k}d_{kk}v_{kj} \{c^{\dagger}_{i \uparrow}\} \{c^{\dagger}_{j \downarrow}\} |0\rangle\\
&= \sum_{k} d_{kk} (\sum_{i \in I} u_{ik} \{c^{\dagger}_{i \uparrow}\} ) (\sum_{j \in J} v_{kj} \{c^{\dagger}_{j \downarrow}\}) |0\rangle \\
&= \sum_{k} \alpha_{k} |k_{\uparrow,A}\rangle |k_{\downarrow,B}\rangle 
\end{align}

Notice that we construct Schmidt vectors with $K/2$ spin-up singlons in $A$ and $K/2$ spin-down singlons in $B$. By using the same contour integral technique as Ref. \onlinecite{Vafek_2016arXiv_eta}, the reduced density matrix has the following form.

\begin{equation}
\begin{split}
\rho^{\text{single}}_{A} &= tr_B [\mathcal{A}_N \eta_+^N |\psi_{K/2} \rangle \langle \psi_{K/2}| \mathcal{A}_N \eta_-^N]\\
&=(\mathcal{A}_N N!)^2 \oint_o \oint_o \frac{dz_1dz^*_2}{(2 \pi)^2}
\sum_{k} \alpha^2_{k} \langle k_{\downarrow,B}| \\ 
& \frac{e^{z^*_2\eta_{-,B}}}{(z^*_2)^{N+1}} \frac{e^{z_1\eta_{+,B}}}{z_1^{N+1}} |k_{\downarrow,B}\rangle   e^{z_1\eta_{+,A}}  |k_{\uparrow,A}\rangle \langle k_{\uparrow,A}|  e^{z^*_2\eta_{-,A}}
\end{split}
\end{equation}

\begin{equation}
\begin{split}
\rho^{\text{single}}_{A} &= (\mathcal{A}_N N!)^2 \oint_o \oint_o \frac{dz_1dz^*_2}{(2 \pi)^2}
\sum_{k} \alpha^2_{k} \frac{(1+z_1 z^*_2)^{L_B-K/2}}{(z_1 z^*_2)^{N+1}}\\
&   e^{z_1\eta_{+,A}}  |k_{\uparrow,A}\rangle \langle k_{\uparrow,A}|  e^{z^*_2\eta_{-,A}}
\end{split}
\end{equation}
where 
\begin{equation}
  \eta_{-,A/B} = \sum_{i \in A/B} (-1)^i c_{i\uparrow} c_{i\downarrow}, \quad   \eta_{+,A/B} = \eta^{\dagger}_{-,A/B}.
\end{equation}

After carrying out the contour integrals, we get
\begin{equation}
\rho^{\text{single}}_{A} = \sum_{k} \alpha^2_{k} \sum_{n} \lambda_{n} |N-n,k_{\uparrow,A}\rangle \langle N-n,k_{\uparrow,A}|
\end{equation}

\begin{equation}
	\lambda_{n} = \frac{C_{L_A-K/2}^{N-n} C_{L_B-K/2}^{n} }{C_{L-K}^N},
\end{equation}

\begin{equation}
	|N-n,k_{\uparrow,A}\rangle = \mathcal{A}_{N-n}^{L_A} \eta^{N-n}_{+,A}  |k_{\uparrow,A}  \rangle.
\end{equation}

where $C^k_n = n!/[k!(n-k)!]$ is the combinatorial number, $\mathcal{A}_{N-n}^{L_A}$ is a normalization constant given by Eq.~\eqref{eq:AN_normalization} by replacing $L$ with the subsystem size $L_A$. It is crucial to choose the maximum polarized sector so that $|k_{\uparrow,A}  \rangle$  is a reference state of $\eta_{+,A}$. 

The entanglement entropy can be calculated as follows
\begin{align}
S &= - \sum_{k,n} \alpha^2_{k} \lambda_{n} \ln (\alpha^2_{k}  \lambda_{n})\\
  &= - \sum_{k,n} \alpha^2_{k} \lambda_{n} \ln (\alpha^2_{k}) -\sum_{k,n} \alpha^2_{k} \lambda_{n} \ln (\lambda_{n})\\
  &= - \sum_{k} \alpha^2_{k} \ln (\alpha^2_{k}) - \beta_{A} \sum_{n}  \lambda_{n} \ln (\lambda_{n})\\
  &= S_{ref} - \beta_{A} \sum_{n}  \lambda_{n} \ln (\lambda_{n}) \label{eq:S_max_polarization}
\end{align}

where $S_{ref}=- \sum_{k} \alpha^2_{k} \ln (\alpha^2_{k})$, 
$\beta_{A}=\sum \alpha_{k}^2 = \sum_{k} d^2_{kk}$. Since the sum of the square of the singular values is equal to the Frobenius norm of the matrix, it follows that $\beta_{A}=\sum_{i \in I,j \in J} \alpha^2_{i,j}$.  Eq.~\eqref{eq:S_max_polarization} is checked numerically with system size $L=8$ and reference electron number $K=2$ in Fig. \ref{fig:S_max_Sz_K2}.

Naturally, we are interested in the limit of highly excited states, large system size and large heat bath size. Therefore we take the limit of $L_B \gg L_A$, $N \gg L_A - K/2$, and $L_A \gg K/2$. In this case, one can simplify $S-S_\text{ref}$ using the Stirling approximation,
replace the summation by an integral, and finally apply a saddle point approximation, which leads to

\begin{equation}
S-S_\text{ref} \approx \frac{\beta_{A}}{2} (1 + \ln[2 \pi \nu (1-\nu)  (L_A-K/2)]) \label{eq:many_ref}
\end{equation} 
where $\nu=N/(L-K)$ indicates the portion of available sites taken by double occupancies. Notice that the symmetric appearance of $\nu (1-\nu)$ with respect to $\frac{1}{2}$ is a consequence of the particle-hole symmetry (see Fig.~\ref{fig:dEE_lnv_1-v}). We can clearly see the logarithmic contribution of $\ln{(L_A-K/2)}$ to the total von Neumann entanglement entropy.

\subsection{Reference states with only singlons}\label{subsec:singlon}

In this subsection, we show that, for any eta-pairing eigenstate in region II of Fig. \ref{fig:EE_Hmap} whose reference state contains only singlons, the entanglement entropy grows exactly logarithmically in the thermodynamic limit.

Given a reference $|\psi_{ref}\rangle$ with only singlons, we decompose it into terms of fixed singlon number $i$ in subsystem $A$.  Notice that the operation of $\eta_+$ only adds doublons to a basis vector and therefore, can't change the value of $i$ except by destroying the state. Since the total singlon number is fixed in the reference state, the singlon number in subsystem $B$ will not change either when $\eta_+$ is applied. When we trace out $B$ to calculate the reduced density matrix of $A$, the terms $|\psi_{t'}\rangle\langle \psi_t|$ where $t\neq t'$ will vanish because $|\psi_{t'}\rangle$ and $| \psi_t\rangle $ have different values in terms of singlon number in subsystem $B$. As a result, the reduced density matrix will be block diagonal according to the singlon number $i$ in subsystem $A$.

Take a many particle reference state in the form of Eq.~\eqref{many_wf} with all $K$ electrons to be singlons. We consider the component $|\psi_i\rangle$ with $i$ singlons in subsystem $A$. Following the same calculation as in the previous subsection, we perform a Schmidt decomposition and have 

\begin{equation}
|\psi_{i} \rangle = \sum_{k} \alpha_{i,k} |k_{i}\rangle |k_{K-i}\rangle 
\end{equation}
where we construct Schmidt vectors of $i$ singlons in $A$ and $K-i$ singlons in $B$.

Using the same contour integral technique, one can show that the reduced density matrix from this reference state has the following form.

\begin{equation}
\rho^{\text{single}}_{i,A} = \sum_{k,n} \alpha^2_{i,k} \lambda_{i,n}  |\phi^A_{i,n}\rangle \langle \phi^A_{i,n}|
\end{equation}
where $\lambda_{i,n}=\frac{C_{L_B-(K-i)}^{n} C_{L_A-i}^{N-n}}{C_{L-K}^N} $ and ${|\phi^A_{i,n}}\rangle$ is a set of orthonormal basis.

It follows that the entanglement entropy has the form.

\begin{equation}
S^{\text{single}}_{i,A} = S_{i,ref} - \beta_{i,A} \sum_{n} \lambda_{i,n} \ln \lambda_{i,n} \label{eq:singlon_EE2}
\end{equation}
where $S_{i,ref}=-\sum_{i,k} \alpha^2_{i,k} \ln \alpha^2_{i,k}$, $\beta_{i,A}=\sum \alpha^2_{i,n} = \sum_{(i_1,\sigma_1), \cdots, (i_K,\sigma_K)}$ $ \alpha^2_{(i_1,\sigma_1), \cdots, (i_K,\sigma_K)}$ with the constraint that there are $i$ singlons in $A$ and $K-i$ singlons in $B$.

Summing up the entanglement entropy contribution from different singlon number $i$ sectors, the total entanglement entropy is  
\begin{equation}
S_{total}=\sum_i S_{i,ref} - \sum_{i,n} \beta_{i,A}  \lambda_{i,n} \ln \lambda_{i,n} \label{eq:singlon_EE}
\end{equation}

Eq.~\eqref{eq:singlon_EE} is consistent with numerical result with system size $L=8$ and reference electron number $K=2$ in Fig.~\ref{fig:S_singlon_K2}.

In the thermodynamic limit of $L_B \gg L_A$, $N \gg L_A - K$, $L_A \gg K$, and $N+K<L$, we can apply Stirling approximation and saddle point approximation to each $S^{\text{single}}_{i,A}$, which implies an additive scaling of $\ln(L_A-i)$ to the entanglement entropy. The result is true for all $i\leq K$ sectors as long as the reference state has only singlon occupancies. Therefore, the total entanglement entropy of the eta-pairing states built on reference states with only singlon occupancies will have logarithmically more entanglement entropy than the reference states.

\section{Numerical Evaluation of Entanglement}
\label{sec:Entanglement_of_reference_states}
In the previous section, we proved that there is, at least, an additive logarithmic increase in entanglement from the reference state in various contexts.  While this rules out that eta-pairing states are area law, these proofs are not sufficient to determine the actual entanglement entropy because we don't know the entanglement of the reference state nor whether the increase in entanglement is greater then logarithmic.  In this section we take steps to answer these questions numerically. 

\begin{figure}[H]
  \centering
  \includegraphics[scale=0.9]{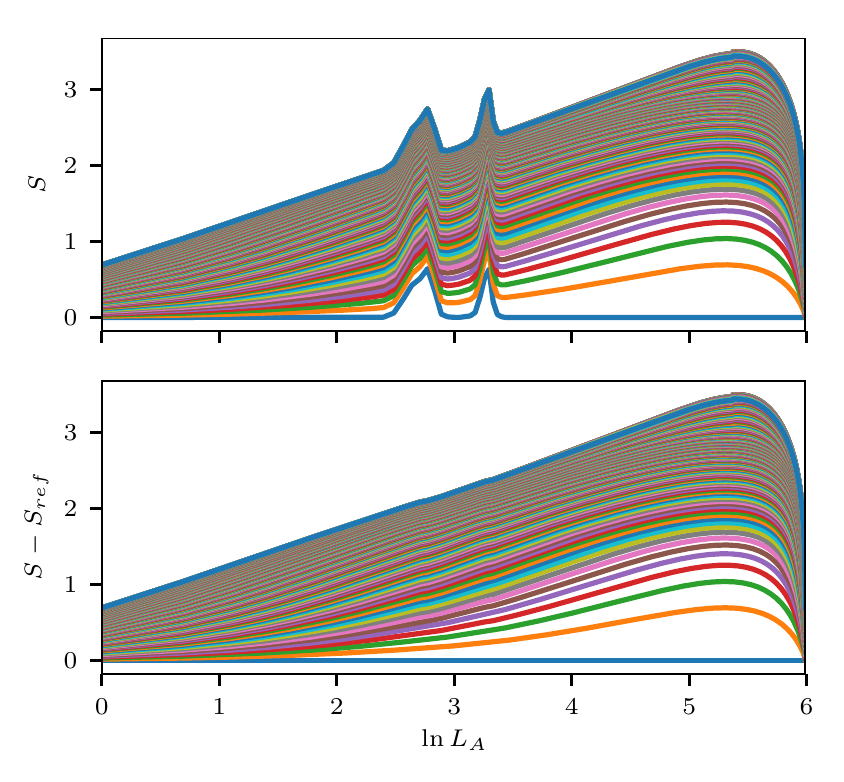}
  \caption{
  \textbf{Top:} Entanglement entropy $S(L_A)$ of $\mathcal{A}_N \eta_+^N|\psi_{ref}\rangle$ for various $N$ (starting at $N=0$ for the bottom curve) with $|\psi_{ref}\rangle$ as a single disordered realization of the open-boundary condition two-particle ground state of Eq.~\eqref{eq:HB_spin} with $W=4,t=1$ and $U=1$. 
  \textbf{Bottom:} Entanglement entropy difference between $\mathcal{A}_N \eta_+^N|\psi_{ref}\rangle$ and $|\psi_{ref}\rangle$ for various $N$.
 }\label{fig:S_eta_pairing_states_sample}
\end{figure}

The simplest reference states we can consider are ground states.  We consider   the ground state of Hamiltonian in Eq.~\eqref{eq:HB_spin} for $W=4$ in the sector where total $S_z=0$ and $K=2$ for $M=400$ total sites.  This state is close to the top of region I in Fig.~\ref{fig:EE_Hmap}.  We apply $\eta_+$ many times and see a clear logarithmic increase in entanglement (see Fig.~\ref{fig:S_eta_pairing_states_sample}). Strictly speaking, under open boundary conditions, pseudo-spin symmetry is no longer exact, but this does not seem to be a problem at large system sizes.   While our proof in section \ref{subsec:max_polarization} does not forbid a faster growth of entanglement, we do not see it in this case.    

\begin{figure}[H]
  \centering
  \includegraphics[scale=0.8]{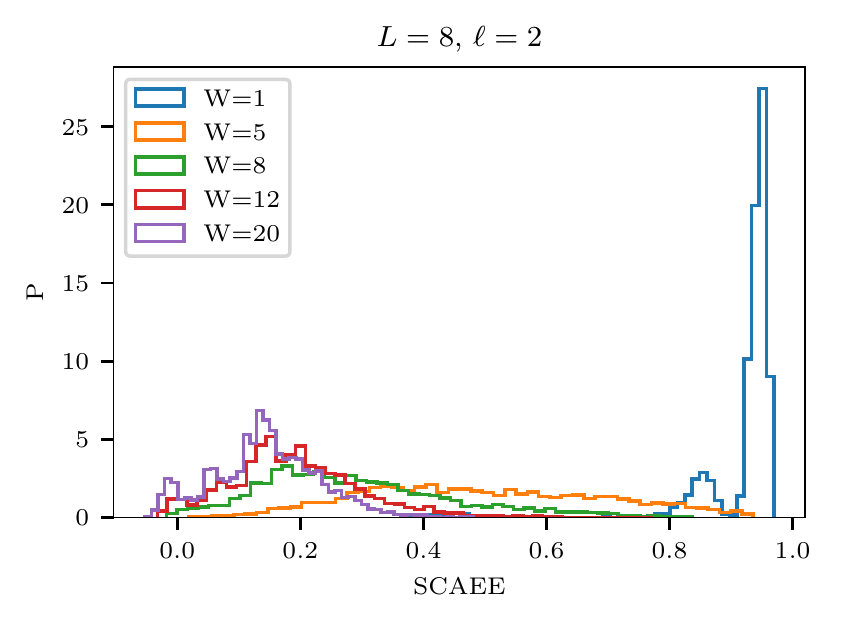}
  \caption{Probability density of SCAEE at $W=1,5,8,12,20$, for $L=8$ and a subsystem size of $\ell=2$. 
  }\label{fig:SCAEE}
\end{figure}

We then consider the entanglement entropy of eigenstates in the middle of the spectrum.  We consider the entanglement of these states for a system of size $L=8$.  For various disorder strengths, we compute the cut averaged entanglement entropy (CAEE) and the slope of the cut-averaged entanglement entropy (SCAEE)~\cite{Yu2016_bimodal}.  Note that the SCAEE equals to 1 at all $L_A$ for an infinite-temperature volume law state and zero for large enough $L_A$ for an area law state.     See Fig.~\ref{fig:SCAEE} for a histogram of these results.  We find the SCAEE consistent with a volume law at small disorder strengths.  At larger disorder strength, though, there is a broad distribution of the SCAEE with some states exhibiting area-law behavior and some states exhibiting sub-volume non-area law behavior. 
Together with the unusual behavior of CAEE in Fig.~\ref{fig:CAEE}, it suggests  a non-ergodic, non-MBL phase in the model.

\begin{figure}[H]
  \centering
  \includegraphics[scale=0.55]{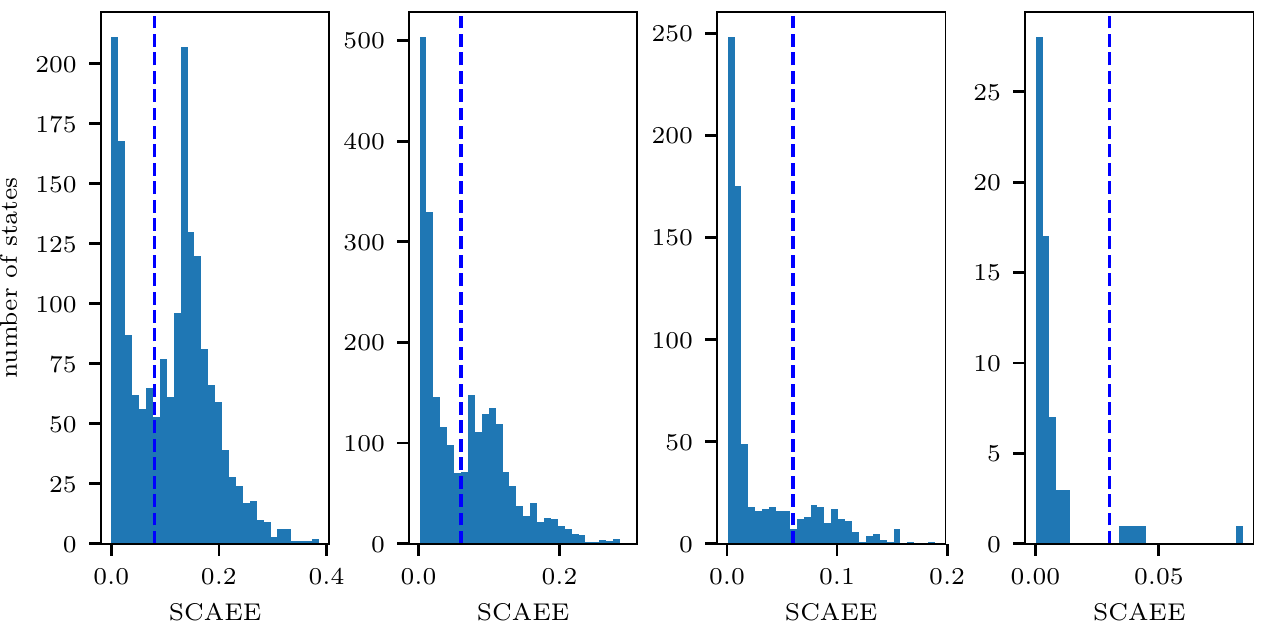}
  \includegraphics[scale=0.6]{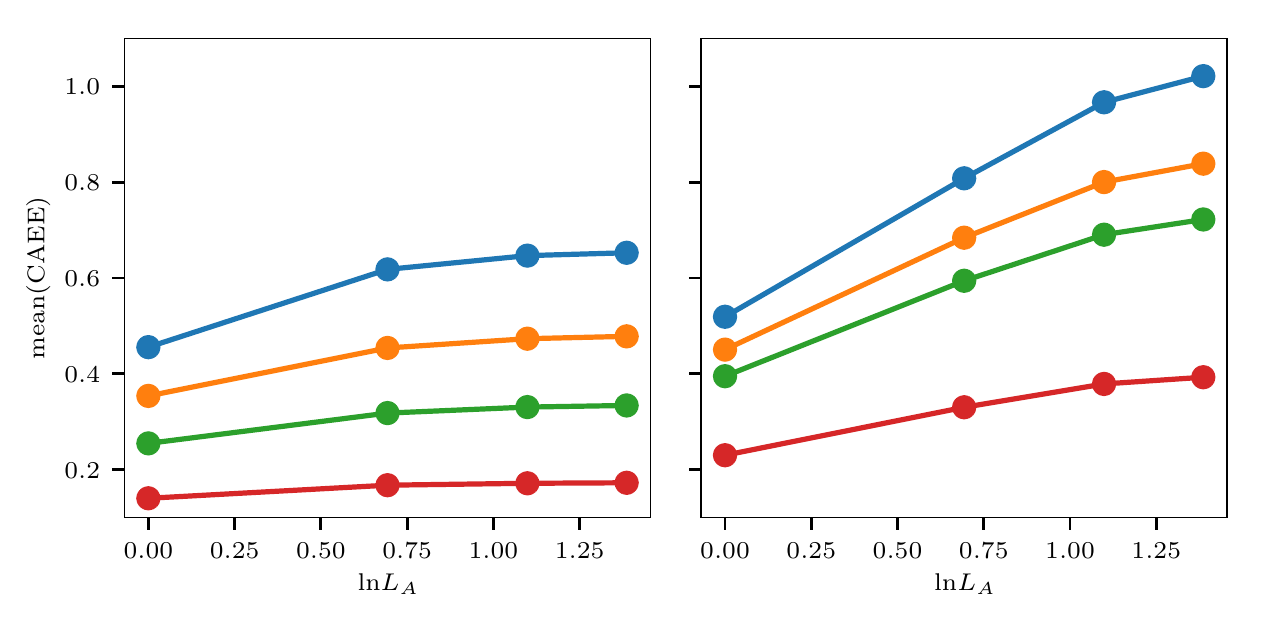}

  \caption{
  \textbf{Top:} The SCAEE histograms at reference states corresponding to (from left to right) $j=0,1,2,3$ with $L=8$, subsystem size $l=2$, $S_z=0$ and $W=14$.  
  \textbf{Bottom:} Mean cut-averaged entanglement entropy (CAEE) vs. ln$L_A$ for different reference states [$j=0$ (\textcolor{blue}{blue curve}), $j=1$ (\textcolor{YellowOrange}{orange curve}), $j=2$ (\textcolor{OliveGreen}{green curve}), $j=3$ (\textcolor{red}{red curve})] for the eigenstates with SCAEE value on the left (right) side of the dashed line corresponding to the left (right) figure. 
  }
  \label{fig:SCAEE_edge_sectors}
\end{figure}

To understand this better, we start by considering the probability density of the SCAEE in reference states of various quantum-number sectors at subsystem size $l=2$ for system size $L=8$ (see Fig.~\ref{fig:SCAEE_edge_sectors}). We find a bimodal distribution with one of the peaks centered at zero and the other peak at a non-zero value much less than 1 again suggesting a mix of area-law and sub-volume law states (see Fig.~\ref{fig:SCAEE_edge_sectors}(bottom), Fig.~\ref{fig:CAEEavg_SCAEE_filter_ln}, Fig.~\ref{fig:CAEEall_SCAEE_filter}). We check this by considering the disordered average entanglement for both peaks.  The peak centered at zero is clearly area-law while the other peak is consistent with logarithmically growing entanglement (See Fig. \ref{fig:SCAEE_edge_sectors}).

\begin{figure}[H]
  \centering
  \includegraphics[scale=0.8]{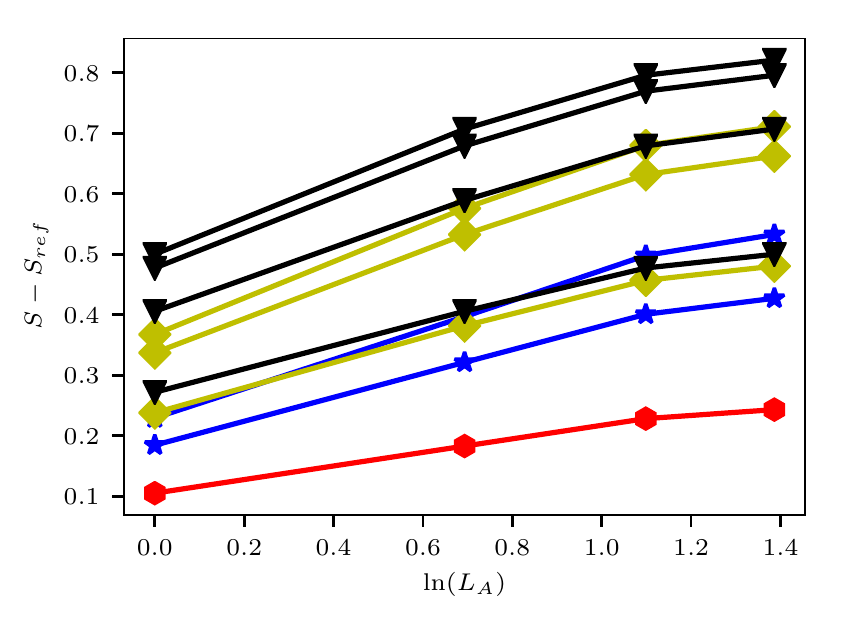}
  \caption{
  The entanglement entropy difference vs. ln$(L_A)$ for $L=8$, $S_z=0$, and $W=14$ at different quantum number sectors[$j=4$ (\textcolor{black}{black curve}), $j=3$ (\textcolor{YellowOrange}{yellow curve}), $j=2$ (\textcolor{blue}{blue curve}), $j=1$ (\textcolor{red}{red curve})]. The entanglement entropy for each quantum number sector is averaged over the entropy of all the eigenstates in the sector obtained from exact diagonalization.  
  }
  \label{fig:dEE}
\end{figure}

We then proceed to consider the difference in entanglement between the reference states and the eta-pairing states. The numerical results (see Fig.~\ref{fig:dEE}) are consistent with all states have a logarithmic increase in entanglement.  This is interesting given that this is in a regime where the proof is not applicable and much of the entanglement comes from sectors other then the maximally polarized. See Fig.~\ref{fig:CAEE06} for the non-disordered average version of this curve.  

\begin{figure}[H]
  \centering
  \includegraphics[scale=0.8]{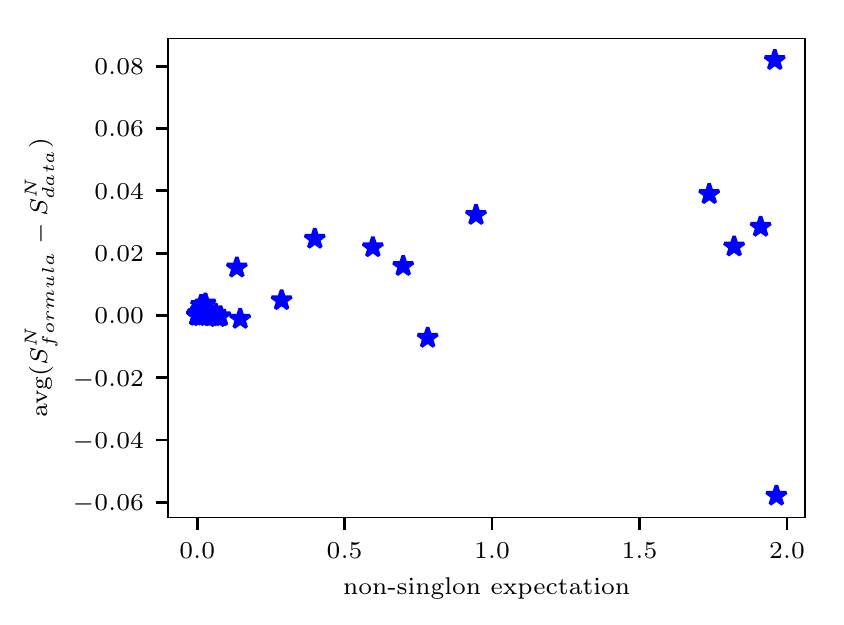}
  \caption{Entanglement entropy difference for $L=8$, $S_z=0$ and $W=14$ between Eq.~\eqref{eq:singlon_EE} and the exact entanglement entropy vs. average non-singlon number in the quantum number sector $j=3,m=0$.  The entanglement entropy difference is averaged over all $L_A$. When the non-singlon number is close to zero, Eq.~\eqref{eq:singlon_EE} provides a good description for the total entanglement entropy.}
  \label{fig:EE_err_Nsavg}
\end{figure}

Finally, we note that for a reference state of only singlons, Eq.~\eqref{eq:singlon_EE}  exactly implies the (logarithmic) increase in entanglement.  While our reference states don't typically have only singlons, we can check the efficacy of this formula as a function of the number of non-singlons in our system.  We find (see Fig.~\ref{fig:EE_err_Nsavg}) that the formula still is applicable with small deviations when the average number of non-singlons is close to zero.

\section{Time evolution after quantum quench} \label{sec:Time_Evolve}
We have given evidence for a Hamiltonian which has both area-law and log-law eigenstates.  Here we show how these different eigenstates can be probed using time-evolution.  In the process this will give further evidence for the two types of states as well as supply a physical picture for why we might expect this difference.

\begin{figure}[h]
  \centering
  \includegraphics[scale=0.4]{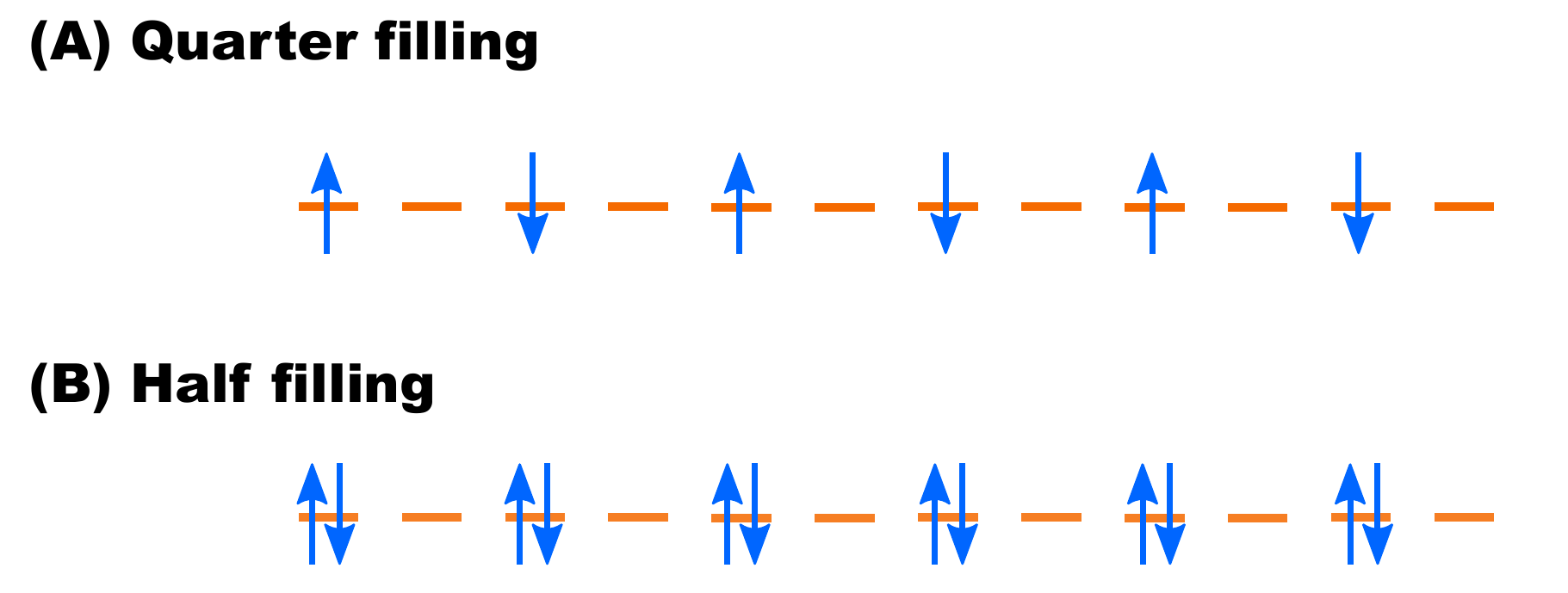}
  \caption{
  \textbf{A:} A schematic diagram for the quarter filled singlon state used in unitary time evolution. 
  \textbf{B:} A schematic diagram for half filled doublon state used in unitary time evolution. 
  Both states have a charge imbalance per electron of 1.}
  \label{fig:Initial_states}
\end{figure}

To accomplish this, our goal will be to find states that are simple to prepare, such as product states, that have overlap with primarily area-law or log-law eigenstates and then consider the effect of time-evolution on these states. We will consider two product states: a quarter filled singlon state, shown in Fig.~\ref{fig:Initial_states}(top),  and a half filled doublon state, shown in Fig.~\ref{fig:Initial_states}(bottom).

\begin{figure}[ht]
  \centering
  \includegraphics[scale=0.8]{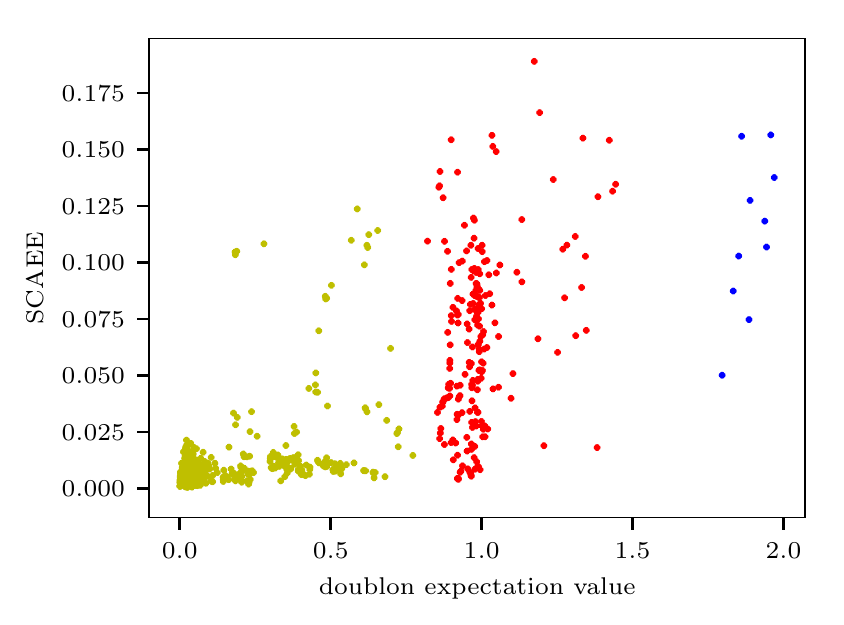}
  \caption{SCAEE vs. doublon expectation value for $L=8$, $l=2$, $S_z=0$ and $W=14$ in quantum number sector $j=2,m=-2$.  For doublon expectation value close to zero, we see area-law entanglement.}
  \label{fig:SCAEE_doublon}
\end{figure}

For a product state $|p\rangle$ (in the occupation basis) the number of single occupancies $n_1$  and  double occupancies $n_2$ fix the expected value of the quantum numbers $(j,m)$,
\begin{equation}
\langle p| \eta_0 | p\rangle = -\frac{(L-n_1-2n_2)}{2}
\end{equation}
and
\begin{equation}
\langle p| \vec{\eta}^2 | p\rangle = \frac{L-n_1-2n_2}{2}\left(\frac{L-n_1-2n_2}{2} + 1\right) + n_2.
\end{equation} \vspace{0.1 cm}\\
\textbf{Quarter filled singlon state: }  To find area-law states we must focus away from states with overlap in region II (which can't be area-law) and instead on those with high overlap with the reference states.  A product state $|p\rangle$ with only single occupancy, is always an 
eigenstate of $\{ \vec{\eta}^2, \eta_0\}$ of eigenvalues $(\frac{L-n_1}{2}, -\frac{L-n_1}{2})$ and therefore is a linear superposition of only reference states in one quantum number sector.  Moreover, of those states, we find (see Fig.~\ref{fig:SCAEE_doublon}) that states with low doublon number are area law states.  An area law state should be `many-body localized' and so we generically expect that time evolution starting in such states should not equilibrate. \vspace{0.2 cm} \\
\textbf{Half filled doublon state: }  On the other hand, to find log-law states, we can look for product states which have high overlap in region II. While the average single and double occupancy doesn't fix the quantum number sector it localizes it around a given quantum number sector. For $n_2 < \frac{L}{2}$, as $n_2$ increases, $\langle \eta_0 \rangle$ grows towards 0, while $\langle \vec{\eta}^2 \rangle$ decreases towards $\frac{L}{2}$. For $n_2 > \frac{L}{2}$, as $n_2$ increases, $\langle \eta_0 \rangle$ approaches $\frac{L}{2}$, while $\langle \vec{\eta}^2 \rangle$ increases towards $\frac{L}{2}(\frac{L}{2} + 1)$. For either case, one can see that the half-filled doublon state is composed of eta-pairing states high up on long pseudo-spin ladders, which have logarithmic corrections to their entanglement entropy.  As a log-law state we expect less localization than a MBL state.  \vspace{0.2 cm} 

\begin{figure}[H]
  \centering
  \includegraphics[scale=0.8]{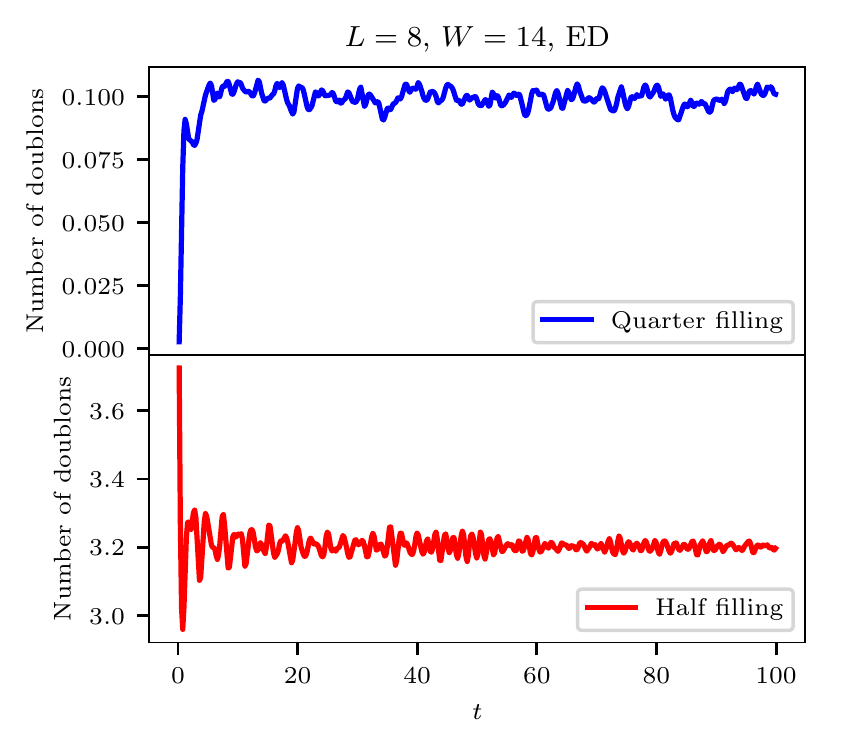}
  \caption{Ensemble averaged number of doublon for $W=14, L=8$ with quarter filled singlon state and half filled doublon states.  Same samples are used as in the first column of Fig.~\ref{fig:compare_S}. The initial doublon number is zero for the quarter filled state and four for the half filled state.}
  \label{fig:ED_doublon}
\end{figure}

Both of the product states start with zero entanglement entropy, and highly imbalanced charge distributions between even and odd sites.  By considering the time evolution of doublon number in both quarter filling and half filling settings, we can verify that the doublon number is largely localized (see Fig.~\ref{fig:ED_doublon}). 

With these two initial product states, we investigate the time evolution of von Neumann entanglement entropy, and charge imbalance
\begin{equation}
I = \frac{\sum_j (-1)^j (n_{j \uparrow}+n_{j \downarrow})}{\sum_j (n_{j \uparrow}+n_{j \downarrow})}.
\end{equation}
We also look at staggered magnetization in Appendix \ref{app:staggered_M}.
The main goal is to see the rate of entanglement entropy growth and  whether the charge imbalance relaxes.

\begin{figure*}
  \centering
  \includegraphics[scale=0.95]{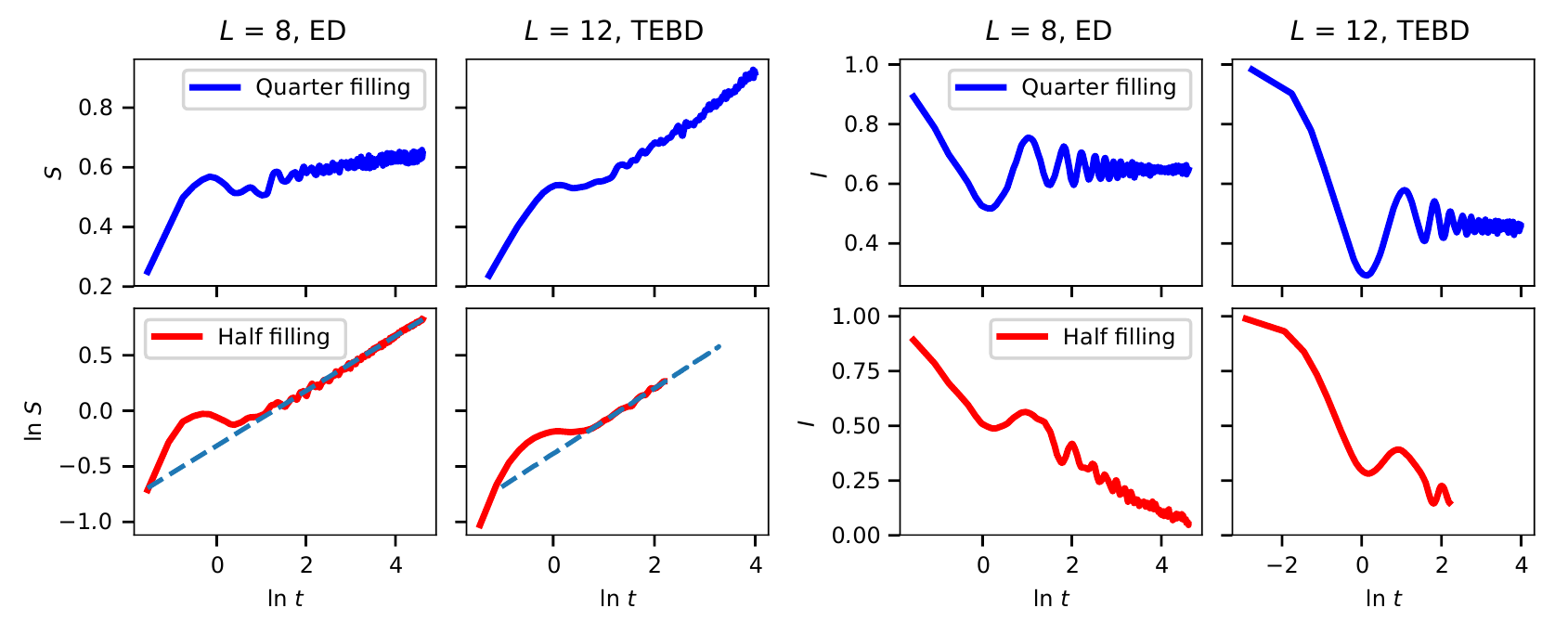}
  \caption{Ensemble averaged von Neumann entropy $S$ for $W=14$ with quarter filled singlon and half filled doublon initial product states. \textbf{First column:} $L=8$, exact diagonalization and periodic boundary conditions. Results are averaged over 400 (top) and 150 (bottom) samples respectively. \textbf{Second column:} $L=12$, TEBD, and open boundary conditions. Results were averaged over 210 samples. For the quarter filling case (\textcolor{blue}{blue curves}), the entropy grows logarithmically with respect to time. For the half filling case (\textcolor{red}{red curves}), the entropy grows as a power law with $t$, with the power law exponent equal to 0.245 for $L=8$, and $0.29$ for $L=12$. \textbf{Third \& Fourth columns:} Ensemble averaged charge imbalance $I$ using the same samples and parameters as the first and the second column respectively. 
 }\label{fig:compare_S}
\end{figure*}

The real time evolution simulations are carried out separately for $L=8$, which uses the exact diagonalization (ED) method, and for $L=12$, which uses the time-evolving block decimation (TEBD) method based on the open source ITensor library \cite{ITensor}. We consider disorder strengths $W=14$ and $W=24$. Under each simulation, the entanglement entropy, charge imbalance and staggered magnetization are averaged over disorder realizations.

We find (see Fig.~\ref{fig:compare_S}) that the quarter filled singlon case exhibits logarithmic growth in entanglement entropy and a charge imbalance that, after an initial decay,  never relaxes stabilizing around a non-zero value.  This is as expected for a many-body localized state.  On the other hand, the half filled doublon case exhibits a power-law growth of entanglement as well as a charge-imbalance which decays quickly to zero.  Although this is suggestive of thermalization, the slope of the entanglement is significantly below the expected linear growth of an ergodic phase.  We attribute this difference to the logarithmic as opposed to volume-law entanglement of the eigenstates. 

There is a simple physical picture consistent with these results.  Since double occupancy has $S_z=0$, spin-up and spin-down electrons can hop together through a second order process, which leads to full charge delocalization in the half filled setting. However, single occupancies can not hop freely due to spin disorder, which prevents full charge delocalization in the quarter filling case. Under the spin disorder potential, the double occupancy tends to hop together and creates charge relaxation. 

From the above analysis, it is clear that the quarter filled singlon product state acts many-body localized while the half filled doublon product state is neither fully ergodic nor MBL. Moreover, via time-evolution we see that we can directly probe the area-law and log-law parts of the spectrum opening up the possibility that this effect can be seen experimentally.

\section{Conclusion}

Within a spin-disordered Hubbard chain at large disorder, we find a number of area-law and log-law eigenstates. Our results are presented in the context of the quantum numbers of the pseudo-spin symmetry of this model (see Fig.~\ref{fig:EE_Hmap}).  Using analytic arguments related to pseudo-spin symmetry, we showed that there is, at least, an additive logarithmic entanglement difference between the states in region I and those in region II.  We present numerical results which suggest that this difference is in fact logarithmic.  Moreover, we show numerical evidence that the states in region II are all log-law while the states in region I are partially area-law and partially log-law with the area-law states being preferentially in states with smaller expected value of doublons.  We then consider two product states which have primary overlap with area-law or log-law eigenstates respectively.  We find that under time evolution the product state consisting of primarily area-law eigenstates acts like a MBL eigenstate with localized charge imbalance and logarithmic growth of entanglement.  On the other hand, the product state consisting of primarily log-law eigenstates has charge imbalance which relaxes and an entanglement which grows polynomially but not linearly.  

While our focus in this work has been on large disorder, from Fig.~\ref{fig:SCAEE} we can see that at small  disorder ($W=1$) this system eventually transitions to an ergodic phase which consists of primarily volume-law eigenstates. Interestingly, at this disorder, there is clear bimodality in the entanglement entropy of eigenstates.  Moreover, we might anticipate that there is a transition around $W=5$ where there is a surprisingly broad spread of entanglement entropies (see Fig.~\ref{fig:SCAEE} and Fig.~\ref{fig:CAEE}).  

Our work provides a solid microscopic Hamiltonian that demonstrates the existence of a non-ergodic, non-MBL phase in one-dimensional system.  Such phases will not have local integrals of motion nor small unitary tensor networks.  This work opens up the possibility of different entanglement structures beyond the area-law of many body localized state in disordered systems.

\section*{Acknowledgment}
This project is part of the Blue Waters sustained-petascale computing project,
which is supported by the National Science Foundation (awards OCI-0725070 and ACI-1238993) and the State of Illinois.
Blue Waters is a joint effort of the University of Illinois at Urbana-Champaign and its National Center for Supercomputing Applications. This material is based upon work supported by the U.S. Department of Energy, Office of Science under Award Number FG02-12ER46875. Di acknowledges useful discussions with Tianci Zhou. BKC thanks Andrei Bernevig for mentioning his eta-pairing work.  
	
\appendix

\section{Reference states}\label{app:reference_state}
Note that a reference state has the property that 
\begin{equation}
\eta_- |\psi_{\text{ref}} \rangle = 0.
\end{equation}
This implies that any eigenstate which has overlap with any product state of only singlons is a reference state. This follows because we know that $\eta_+ \eta_-$ acted on a non-reference eigenstate gives back the eigenstate and since $\eta_\pm$ only affects doublons, any single-occupancy-only configuration will be annihilated by $\eta_-$. 

For a $K$-electron eigenstate  $|\psi_{\text{ref}} \rangle$  which can be annihilated by $\eta_-$, we have
\begin{equation}\label{eq:eta_0_qn}
\eta_0 |\psi_{\text{ref}} \rangle = -\frac{L-K}{2}|\psi_{\text{ref}} \rangle,
\end{equation}

Because of Eq.~\eqref{eq:eta_algebra}, it then follows that 
\begin{equation}
\vec{\eta}^2 |\psi_{\text{ref}} \rangle = (\eta_+ \eta_- - \eta_0 + \eta_0^2)|\psi_{\text{ref}} \rangle,
\end{equation}
which can be reduced to
\begin{equation}\label{eq:eta_square_qn}
\vec{\eta}^2 |\psi_{\text{ref}} \rangle = \frac{L-K}{2} \left(\frac{L-K}{2} + 1\right) |\psi_{\text{ref}} \rangle,
\end{equation}
meaning that $|\psi_{\text{ref}}\rangle$ has pseudo-spin quantum number of $(\frac{L-K}{2}, -\frac{L-K}{2})$ for $(\vec{\eta}^2 ,\eta_0)$, which then can be raised by $\eta_+$ for at most $(L-K)$ times.

\section{Normalization factor of eta-pairing states built from many particle reference state} \label{app:many_particle_reference_state}
Assume that we have a $K$-particle eigenstate of the spin-disordered Hubbard model as
\begin{equation}\label{many_wf}
  |\psi_{\text{ref}}\rangle = \sum_{(i_1,\sigma_1), \cdots, (i_K,\sigma_K)} \alpha_{(i_1,\sigma_1), \cdots, (i_K,\sigma_K)} c^{\dagger}_{i_1,\sigma_1} \cdots c^{\dagger}_{i_K,\sigma_K} |0\rangle,
\end{equation}
which satisfies $K \ll L/2$ and $\eta_- |\psi_{\text{ref}}\rangle = 0$

On top of this eigenstate, one can also build a tower of eta-pairing states as
\begin{equation}
|\psi^N\rangle =  \mathcal{A}_N \eta_+^N \hspace{-0.4cm} \sum_{(i_1,\sigma_1), \cdots, (i_K,\sigma_K)} \hspace{-0.6cm} \alpha_{(i_1,\sigma_1) \cdots (i_K,\sigma_K)} 
c^{\dagger}_{i_1,\sigma_1} \cdots c^{\dagger}_{i_K,\sigma_K} |0\rangle. 
\end{equation}
But now it becomes too complicated to calculate the normalization factor directly.

To move forward, one should consider Eq.~\eqref{eq:eta_algebra}, \eqref{eq:eta_0_qn}, and \eqref{eq:eta_square_qn}. It is clear that
\begin{equation}
\eta_- \eta_+ |\psi_{\text{ref}}\rangle = (L-K) |\psi_{\text{ref}}\rangle,
\end{equation}
and
\begin{equation}
\eta^2_- \eta^2_+ |\psi_{\text{ref}}\rangle = 2(L-K)(L-K-1) |\psi_{\text{ref}}\rangle,
\end{equation}
Using mathematical induction, is quite easy to show that
\begin{equation}
  \eta^N_- \eta^N_+ |\psi_{\text{ref}}\rangle = C^N_{L-K} (N!)^2 |\psi_{\text{ref}}\rangle,
\end{equation}
which means that
\begin{equation}
  \frac{1}{\mathcal{A}^2_N} = \langle \psi_{\text{ref}} | \eta^N_- \eta^N_+ |\psi_{\text{ref}}\rangle  =  C^N_{L-K} (N!)^2.
\end{equation}
So we finally arrive at
\begin{equation}
  \mathcal{A}_N=\sqrt{\frac{(L-N-K)!}{(L-K)! N!}}.
  \label{eq:AN_normalization}
\end{equation}

\section{Additional numerical evidence on eta-pairing state entanglement entropy}

\begin{figure}[H]
  \centering
  \includegraphics[scale=0.95]{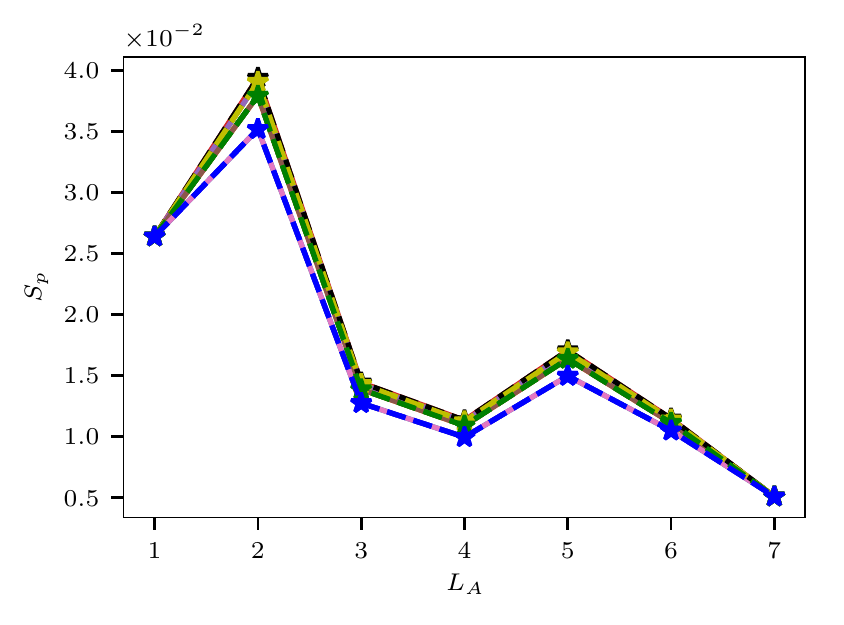}
  \caption{
  Entanglement entropy from maximum polarized sector vs. subsystem size $L_A$ for eta-pairing state with  $\eta_+^N$ [$N=0$ (\textcolor{blue}{blue curve}), $N=1$ (\textcolor{OliveGreen}{green curve}),$N=2$ (\textcolor{YellowOrange}{orange curve}),  $N=3$ (\textcolor{black}{black curve})]. The numerical data is indicated by $*$ and is consistent with the lines generated by Eq.~\eqref{eq:S_max_polarization}. The sample is selected from reference state with average singlon number equal to 0.036, which is far away from the reference state electron number $K=2$. System size $L=8$, polarization $S_z=0$ and disorder strength $W=14$. 
  }
  \label{fig:S_max_Sz_K2}
\end{figure}

\begin{figure}[H]
  \centering
  \includegraphics[scale=0.91]{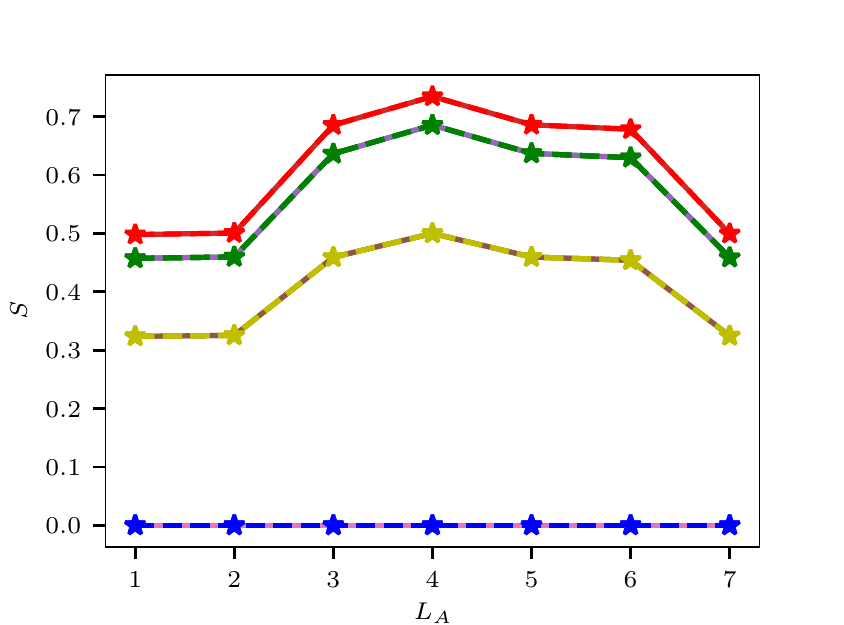}
  \caption{
  Entanglement entropy vs. subsystem size $L_A$ for eta-pairing with $\eta_+^N$ [$N=0$ (\textcolor{blue}{blue curve}), $N=1$ (\textcolor{YellowOrange}{orange curve}), $N=2$ (\textcolor{OliveGreen}{green curve}), $N=3$ (\textcolor{red}{red curve})]. The numerical data is indicated by $*$ and it is consistent with the lines generated by Eq.~\eqref{eq:singlon_EE}. The sample is selected from reference state with average singlon number almost equal to the reference state electron number $K=2$. System size $L=8$, polarization $S_z=0$ and disorder strength $W=14$. 
  }
  \label{fig:S_singlon_K2}
\end{figure}

\begin{figure}[H]
  \centering
  \includegraphics[scale=0.93]{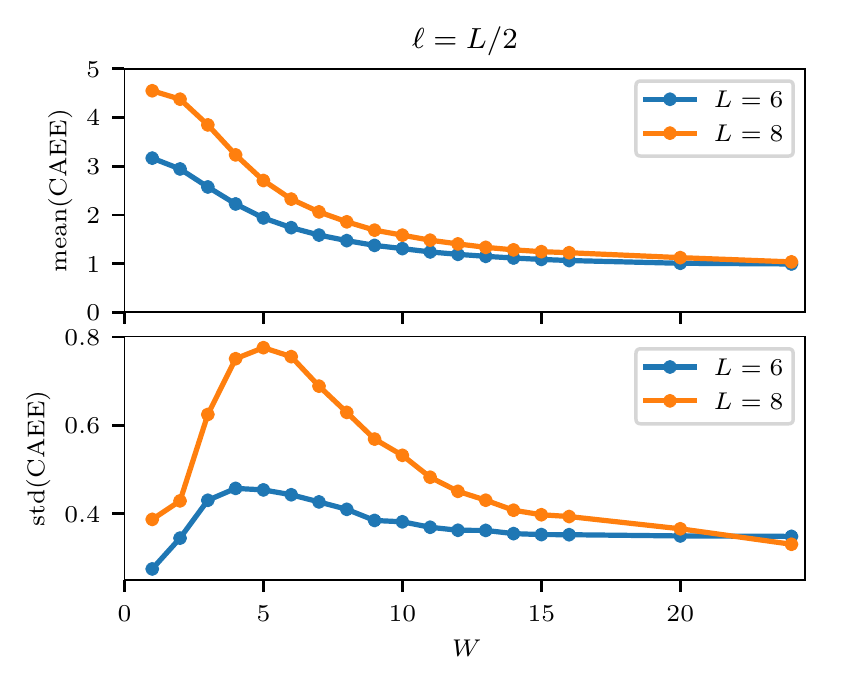}
  \caption{Mean (\textbf{top}) and standard deviation (\textbf{bottom}) of the half-cut CAEE vs. the disorder strength, for $L=6,8$, with 100 disorder samples for each cases, and 30 to 50 eigenstates per sample at the middle of the spectrum in the half filling, $S^z_\textrm{total}=0$ sector. The standard deviation of CAEE is more peaked at $L=8$. Two noticeable differences from the normal MBL transition can be observed. (1) At very large $W$, CAEE stays around 1, instead of zero. (2) For both the mean and the standard deviation of CAEE curves, the crossing point between different system sizes does not show up until very large $W$. These could be signs of a non-ergodic, non-MBL phase.}\label{fig:CAEE}
\end{figure}

\begin{figure}[H]
  \centering
  \includegraphics[scale=0.93]{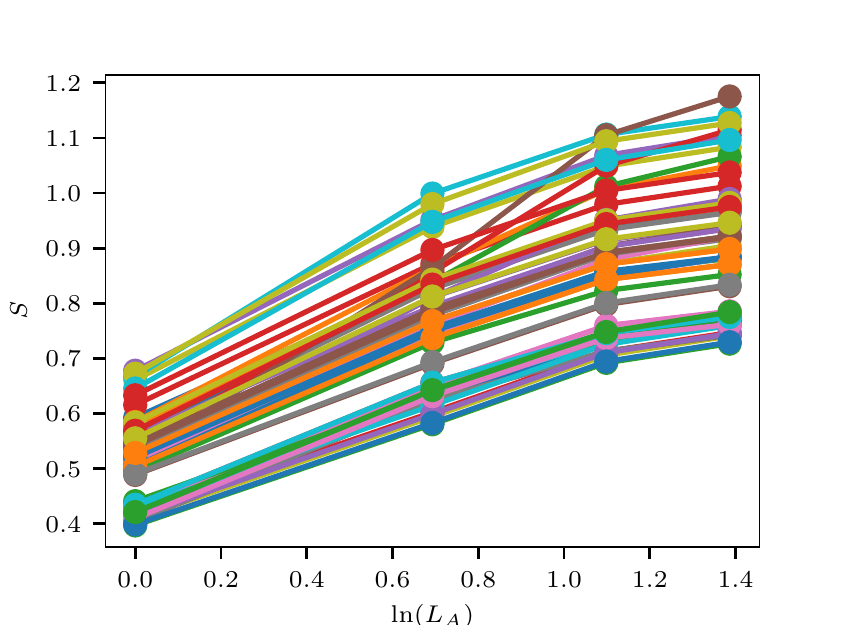}
  \caption{
  The entanglement entropy difference vs. ln$(L_A)$ for all states in the quantum number sector $j=3,m=0$ for $L=8$, $S_z=0$ and $W=14$. The entanglement entropy for each quantum number sector is averaged over the entropy of all the eigenstates in that sector.  
  }
  \label{fig:CAEE06}
\end{figure}

\begin{figure}[H]
  \centering
  \includegraphics[scale=0.85]{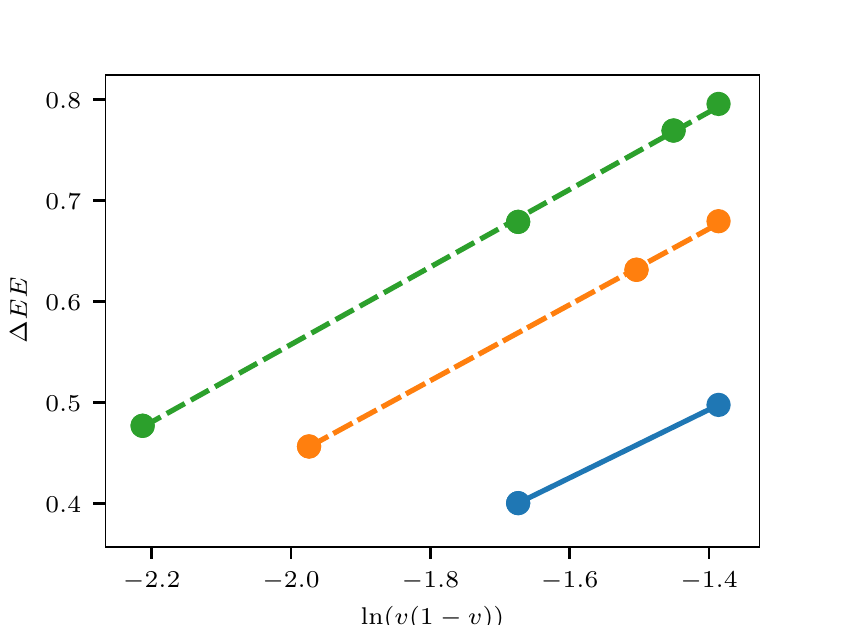}
  \caption{
  Entanglement entropy difference between different quantum number sectors and their reference states [$j=4$ (\textcolor{blue}{blue curve}), $j=3$ (\textcolor{OliveGreen}{green curve}), $j=2$ (\textcolor{red}{red curve})] vs. ln$[\nu(1-\nu)]$. $\nu$ is defined as $N/(L-K)$, where $N$ is the number of times of applying the $\eta_+$ operators on the reference state and $K$ is the number of electrons in the reference state. The entanglement entropy is linear with respect to ln$[\nu(1-\nu)]$, which agrees with Eq.~\eqref{eq:many_ref}. 
  }
  \label{fig:dEE_lnv_1-v}
\end{figure}

Eq.~\eqref{eq:many_ref} indicates that for fixed $L_A$ in the limit $N \gg L_A -K \gg 0$, the entanglement in the single-occupancy reference state sector should increase logarithmically with $\nu(\nu-1)$.  In spite of not being in this regime, for the $L=8$ case we interestingly find that the entanglement entropy of the states in the same quantum number sector $j$ increases logarithmically in $\nu(1-\nu)$.

\section{Bimodal distribution of reference states entanglement entropy}

\begin{figure}[H]
  \centering
   \includegraphics[scale=0.65]{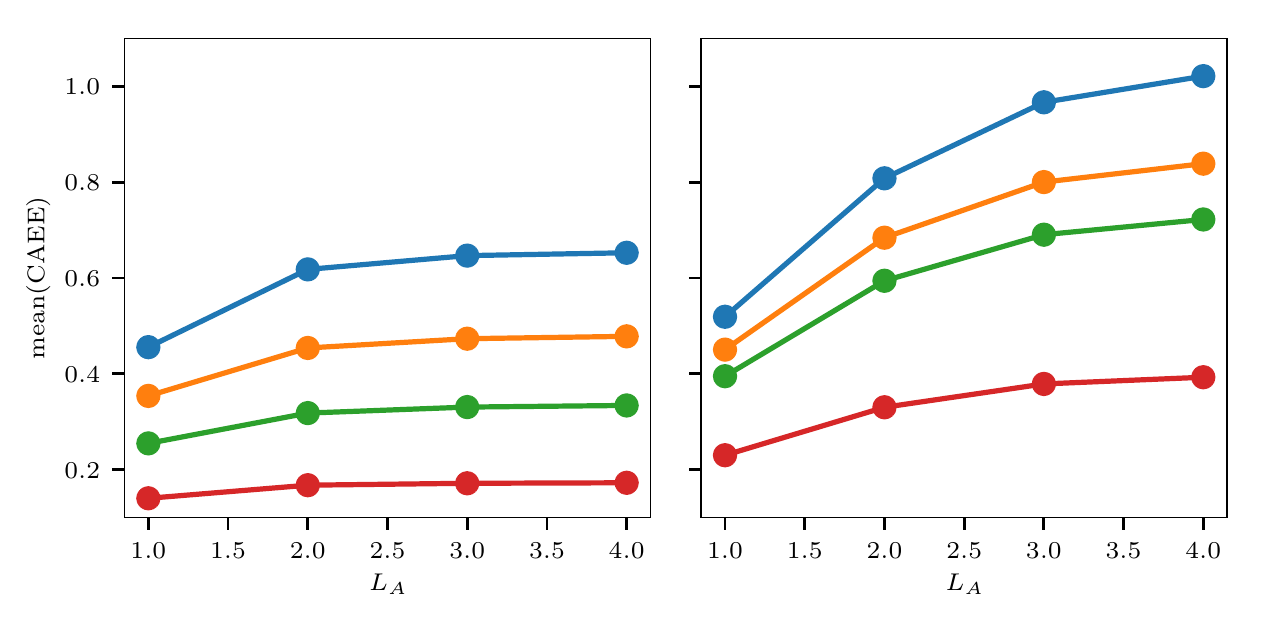}
  \caption{Mean cut-averaged entanglement entropy (CAEE) vs. $L_A$ for different reference states [$j=0$ (\textcolor{blue}{blue curve}), $j=1$ (\textcolor{YellowOrange}{orange curve}), $j=2$ (\textcolor{OliveGreen}{green curve}), $j=3$ (\textcolor{red}{red curve})]. $L_A$ is the subsystem size. System size $L=8$, polarization $S_z=0$ and disorder strength $W=14$. \textbf{Left:} The CAEE in each reference state sector is averaged over the eigenstates with SCAEE value on the left hand side of the dashed line in Fig.~\ref{fig:SCAEE_edge_sectors}. The mean CAEE in this case indicates a area law. \textbf{Right:} The CAEE in each reference state sector is averaged over the eigenstates with SCAEE value on the right hand side of the dashed line in Fig.~\ref{fig:SCAEE_edge_sectors}. The mean CAEE in this case indicates an sub-volume law.
  }
  \label{fig:CAEEavg_SCAEE_filter_ln}
\end{figure}

\begin{figure}[H]
  \centering
  \includegraphics[scale=0.6]{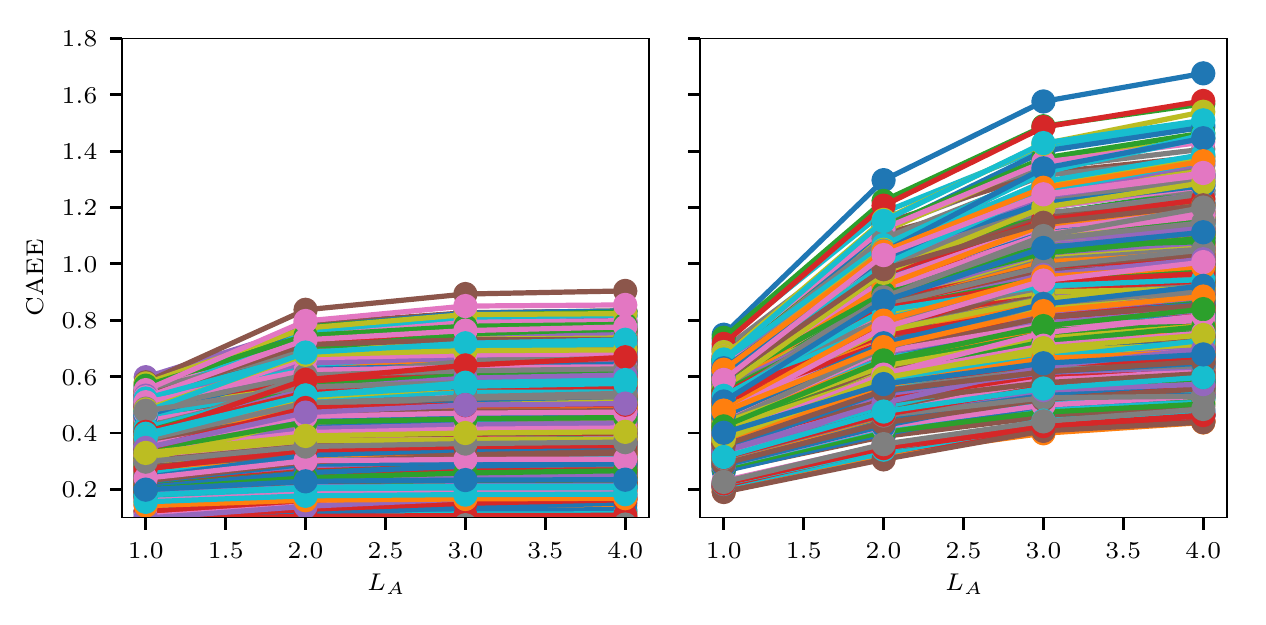}
  \caption{Cut-averaged entanglement entropy (CAEE) vs. $L_A$ for reference state sector at $j=1$. $L_A$ is the subsystem size. System size $L=8$, polarization $S_z=0$ and disorder strength $W=14$. \textbf{Left:} The CAEE in each reference state sector is averaged over the eigenstates with SCAEE value on the left hand side of the dashed line in Fig.~\ref{fig:SCAEE_edge_sectors}. The CAEE in this case indicates a area law. \textbf{Right:} The CAEE in each reference state sector is averaged over the eigenstates with SCAEE value on the right hand side of the dashed line in Fig.~\ref{fig:SCAEE_edge_sectors}. The CAEE in this case indicates an sub-volume law.
  }
  \label{fig:CAEEall_SCAEE_filter}
\end{figure}

\section{Staggered magnetization}\label{app:staggered_M}

We also look at the time evolution of the staggered magnetization.

\begin{equation}
M_{s} = \frac{\sum_j (-1)^{j} (n_{j \uparrow} - n_{j \downarrow})}{\sum_j (n_{j \uparrow}+n_{j \downarrow})}.
\end{equation}

\begin{figure}[H]
  \centering
  \includegraphics[scale=0.5]{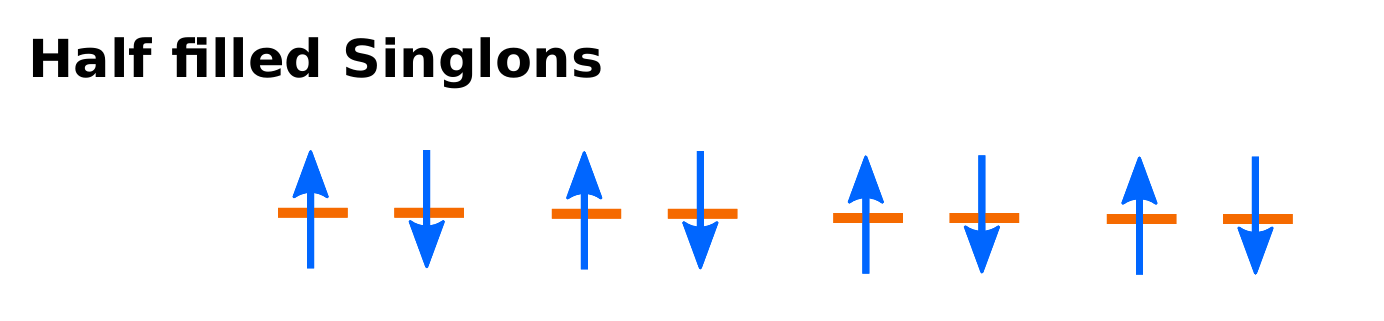} \! \! \!
  \includegraphics[scale=0.85]{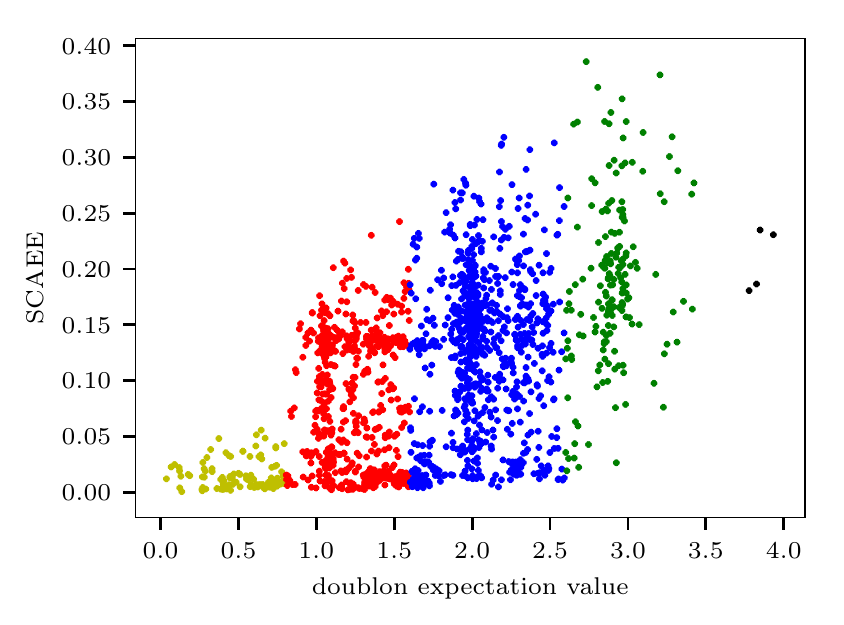}
  \includegraphics[scale=0.85]
  {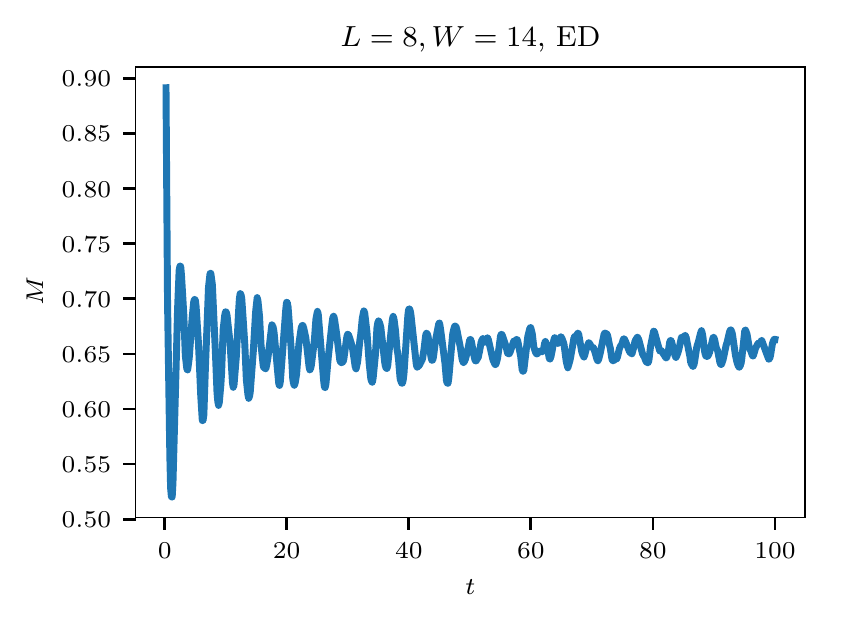}
  \caption{
  \textbf{Top:} Schematic diagram for the half filled singlon state. \textbf{Middle:} SCAEE vs. doublon expectation value for $L=8$, $l=2$, $S_z=0$ and $W=14$ in quantum number sector $j=0,m=0$. 
  \textbf{Bottom:} Disorder averaged (over 200 samples) time evolution of staggered magnetization $M_s$ of the half filled singlon state for $L=8, W=14$.  The initial staggered magnetization is 1. }
  \label{fig:QF_S_I}
\end{figure}

We have chosen the initial state to be the half filled singlon state, whose staggered magnetization starts from 1 (See Fig.~\ref{fig:QF_S_I}(Top)). Since the initial state has no doublon, it is an area-law state (see Fig.~\ref{fig:QF_S_I}(Middle)). Due to the large spin disorder, the spin degree of freedom is localized and the final staggered magnetization stabilizes at a value far away from zero (See Fig.~\ref{fig:QF_S_I}(Bottom)).

\bibliography{EtaPairing}
 
\end{document}